\documentclass[journal, a4paper]{STtemplate}
\usepackage{graphicx}

\begin{document}
\title{A System for In-Ear Pulse Wave Measurements}

\author{S.~Kaufmann, 
		A.~Malhotra, 
		G.~Ardelt, 
		N.~Hunsche, 
		K.~Bre\ss lein,
		R.~Kusche,
		and~M.~Ryschka
		
		\thanks{S. Kaufmann is with the Laboratory for Medical Electronics, L\"ubeck University of Applied Sciences and the Graduate School for Medical Engineering, University of L\"ubeck, Germany (telephone:+ 49 451 300 5400, e-mail: kaufmann@fh-luebeck.de).}
		
		\thanks{A. Malhotra is with the Laboratory for Medical Electronics, L\"ubeck University of Applied Sciences (telephone:+ 49 451 300 5400, e-mail: ankit.malhotra.bm@gmail.com).}
		
		\thanks{G. Ardelt is with the Laboratory for Medical Electronics, L\"ubeck University of Applied Sciences (telephone:+ 49 451 300 5352, e-mail: gunther.ardelt@fh-luebeck.de).}
		
		\thanks{N. Hunsche is with the Laboratory for Medical Electronics, L\"ubeck University of Applied Sciences (telephone:+ 49 451 300 5387, e-mail: nele.hunsche@stud.fh-luebeck.de).}

		\thanks{K. Bre\ss lein is with the Laboratory for Medical Electronics, L\"ubeck University of Applied Sciences (telephone:+ 49 451 300 5400, e-mail: Nabbi619@gmx.de).}

		\thanks{R. Kusche is with the Laboratory for Medical Electronics, L\"ubeck University of Applied Sciences (telephone:+ 49 451 300 5400, e-mail: romankusche@gmx.de).}
		
		\thanks{M. Ryschka is with the Laboratory for Medical Electronics, L\"ubeck University of Applied Sciences (telephone:+ 49 451 300 5026, e-mail: ryschka@fh-luebeck.de).}
}

\maketitle
\pagestyle{empty}
\thispagestyle{empty}

\begin{abstract}
	The measurement of the pulse wave has proven to be a vital tool in medical diagnosis. Whereby most pulse wave measurements are carried out at extremities, this work proposes a system for measuring the pulse wave and the Pulse Arrival Time (PAT) in the interior of the ear. The developed measurement device is based on a battery powered microcontroller system. The measurement device can simultaneously acquire a single channel Electrocardiogram (ECG), a dual wavelengths Photoplethysmogram (PPG), the pressure in both ears, the body core temperature, as well the subjects motion. The acquired measurement data can either be saved on a micro SD-card or can be transmitted wireless via Bluetooth or wired via USB to a host PC for further analysis. In Bluetooth communication mode the device can operate battery powered up to 8 hours. In addition to the system description, first measurements carried out with the system will be presented.
\end{abstract}

\section{Introduction}
A key to future personal monitoring is portability and adaptability. For that purpose it is desirable to record vital signs non-invasively and comfortably for a long period of time.

Pulse Arrival Time (PAT) and the morphology of the pulse wave are considered as indicators of arterial stiffness and are also known as (prognostic) markers for cardiac and vascular diseases \cite{Laurent:Aorticstiffnessisanindependentpredictorofallcauseandcardiovascularmortalityinhypertensivepatients, Guerin:Impactofaorticstiffnessattenuationonsurvivalofpatientsinend-stagerenalfailure}. In the past various methods and sites have been demonstrated for pulse wave measurements, mostly on carotid and femoral artery through invasive and non-invasive methods.

This work proposes a system for detecting the pulse wave inside the auditory canal for PAT and Pulse Wave Velocity (PWV) measurements, as proposed by \cite{kaufmann:Inearpulsewavemeasurements}. The developed measurement device is based on a battery powered microcontroller system and can simultaneously acquire a single channel Electrocardiogram (ECG), a dual wavelengths Photoplethysmogram (PPG), the pressure in both ears, the body core temperature, as well the subjects motion. After data acquisition the measurement data can be transmitted wireless via Bluetooth or wired via USB to a host PC for further analysis. Additionally it is also possible to save the measurement data on a micro SD-card.

\section{MEASUREMENT METHODS}
\subsection{Electrocardiogram (ECG)}
Via the recording of the Electrocardiogram (ECG) it is possible to detect the electrical activity of the heart, which is closely correlated with the contractions of the heart atria and ventricles and can therefore provide a reliable time reference for the PPG and pressure measurements. 

\subsection{Photoplethysmography (PPG), Pulse Arrival Time (PAT) and Pulse Wave Velocity (PWV)}
Caused by the rhythmic contraction of the heart, blood is pumped in a pulsation through the arterial system. The blood pulsation is time and site depended and can be measured in terms of flow, pressure and volume. Via the PPG this volume changes can be measured non-invasively by exploiting the different optical properties of oxygenated and non-oxygenated blood. The morphology of the obtained waveform, as well as the PAT are very significant and direct prognostic markers for the stiffness of the arteries. Hence providing information regarding the overall condition of the cardiovascular system. PAT values can also be used for the calculation of the PWV, which is an additional important marker for arterial stiffness \cite{Expertconsensusdocumentonarterialstiffnessmethodologicalissuesandclinicalapplications, Laurent:Aorticstiffnessisanindependentpredictorofallcauseandcardiovascularmortalityinhypertensivepatients}. 

\begin{figure}[!h]
	\centering
	\includegraphics[width=0.39\textwidth]{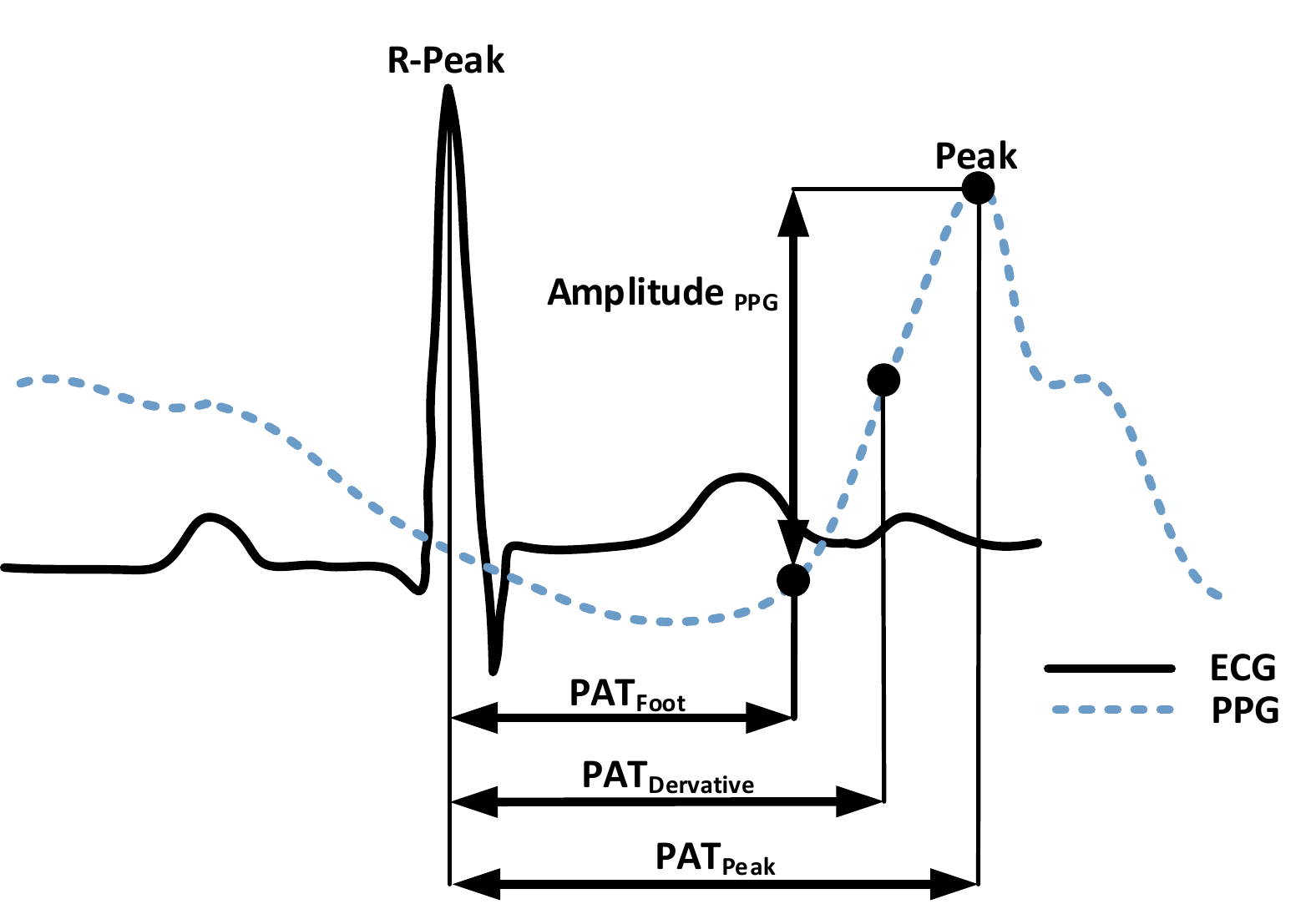}
	\caption{Calculation principle for the Pulse Arrival Time (PAT).}
	\label{fig:PAT}
\end{figure}

The PAT is defined as the time span between the R-peak of the ECG and the arrival of the pulse wave in the PPG signal. The PAT is commonly defined in one of three possibility according to Fig. \ref{fig:PAT}.

The PWV (\ref{equ:PWV}) is given by the ratio between the distance of the two measurement sites ($\Delta x$) and the Pulse Transient Time (PTT) \cite{Federico:NoninvasiveCufflessEstimation}.
\begin{equation}
	PWV = \frac{\Delta x}{PTT} \label{equ:PWV} 
\end{equation}
The PTT is the time required by the pulse wave to travel through the whole segment. The PTT can also be expressed as relative measurement through the PAT and the Pre-Ejection Period (PEP) according to (\ref{equ:PTT}). Whereby the PEP is the time between the R-peak of the ECG and the opening of the aortic valve.
\begin{equation}
	PTT = PAT - PEP \label{equ:PTT}
\end{equation}

\subsection{In-Ear Pressure}
\begin{figure}[!h]
	\centering
	\includegraphics[width=0.44\textwidth]{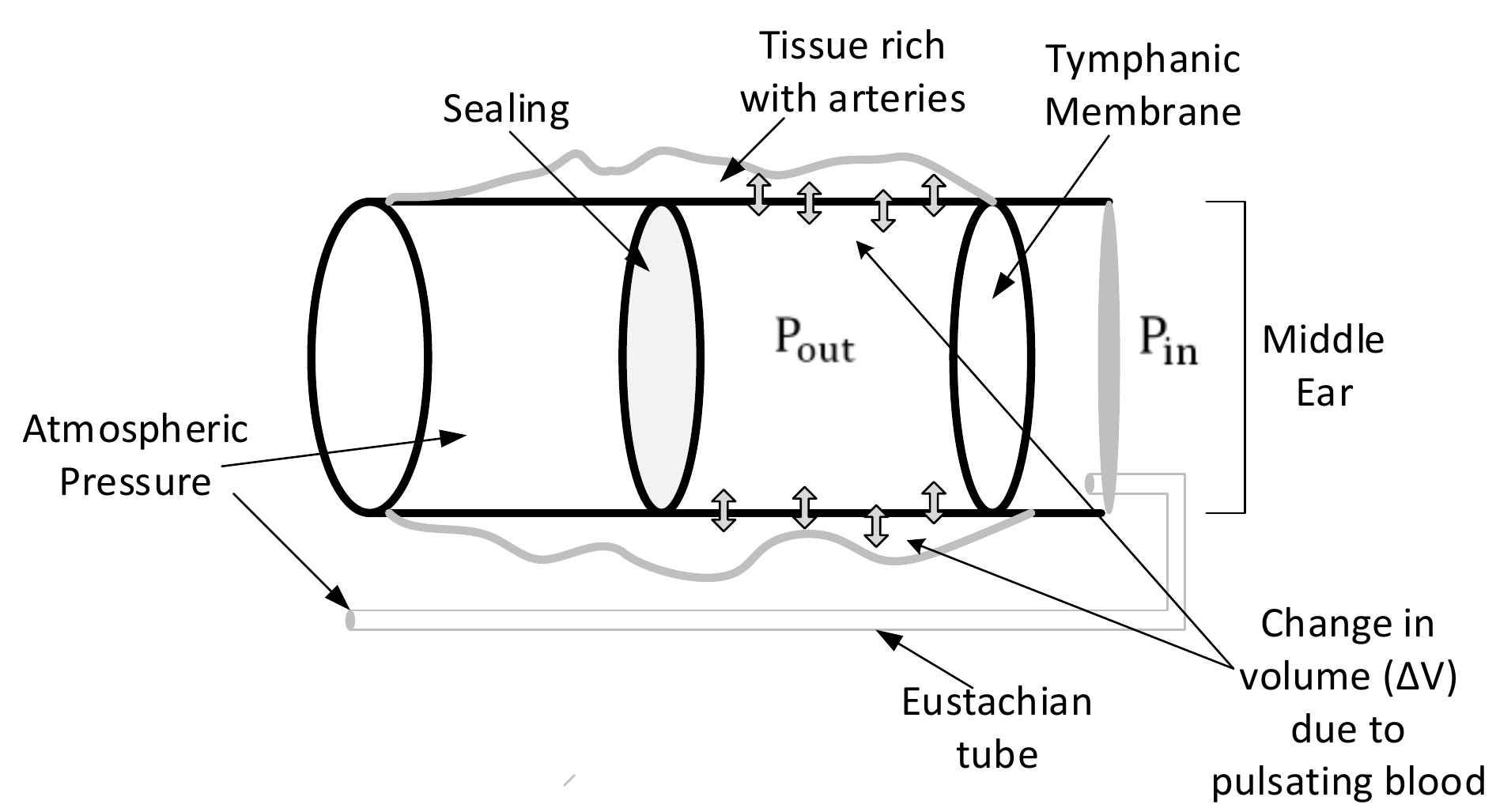}
	\caption{Basic model of the pressure changes in the auditory canal.}
	\label{fig:InEarModel}
\end{figure}

\begin{figure}[!h]
	\centering
	\includegraphics[width=0.42\textwidth]{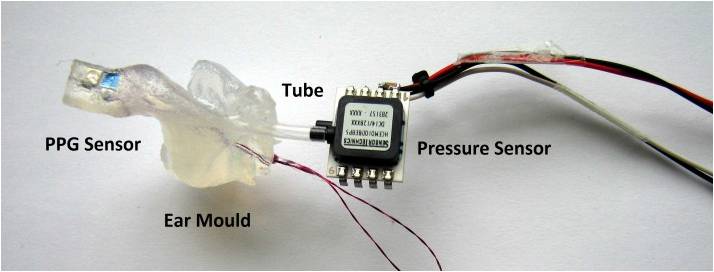}
	\caption{Photograph of the preliminary in-ear sensor comprising of the PPG and a pressure sensor.}
	\label{fig:PPGSensor}
\end{figure}

To be able to measure the internal pressure variation inside the auditory canals, the canals have to be sealed. Based on the ideal gas law $PV = nRT$ (with the pressure $P$, the volume $V$, amount of the substance $n$, the gas constant $R$ and the absolute temperature $T$) a change in volume in a sealed cavity under the assumption of a constant temperature leads to a change in pressure. Fig. \ref{fig:InEarModel} shows a principle drawing of the in-ear pressure model of the auditory canal.
The basic idea is that the arteries around the auditory canal expand based on the heart activity leading to a blood volume change. This change of volume can than be measured with a pressure sensor applied to the sealing. Fig. \ref{fig:PPGSensor} shows a photograph of the developed custom in-ear sensor.
The sensor buildup achieved the required sealing through the mould, while allowing the measurement of a PPG, as well as the pressure inside the ear.

\subsection{Acceleration and Temperature}
For the assessment of the subjects position and physical activity an acceleration sensor is implemented in the measurement system. Additionally the core temperature can be measured via a NTC based temperature sensor.

\section{IMPLEMENTATION}
\subsection{Block diagram}
\begin{figure}[!h]
	\centering
	\includegraphics[width=0.49\textwidth]{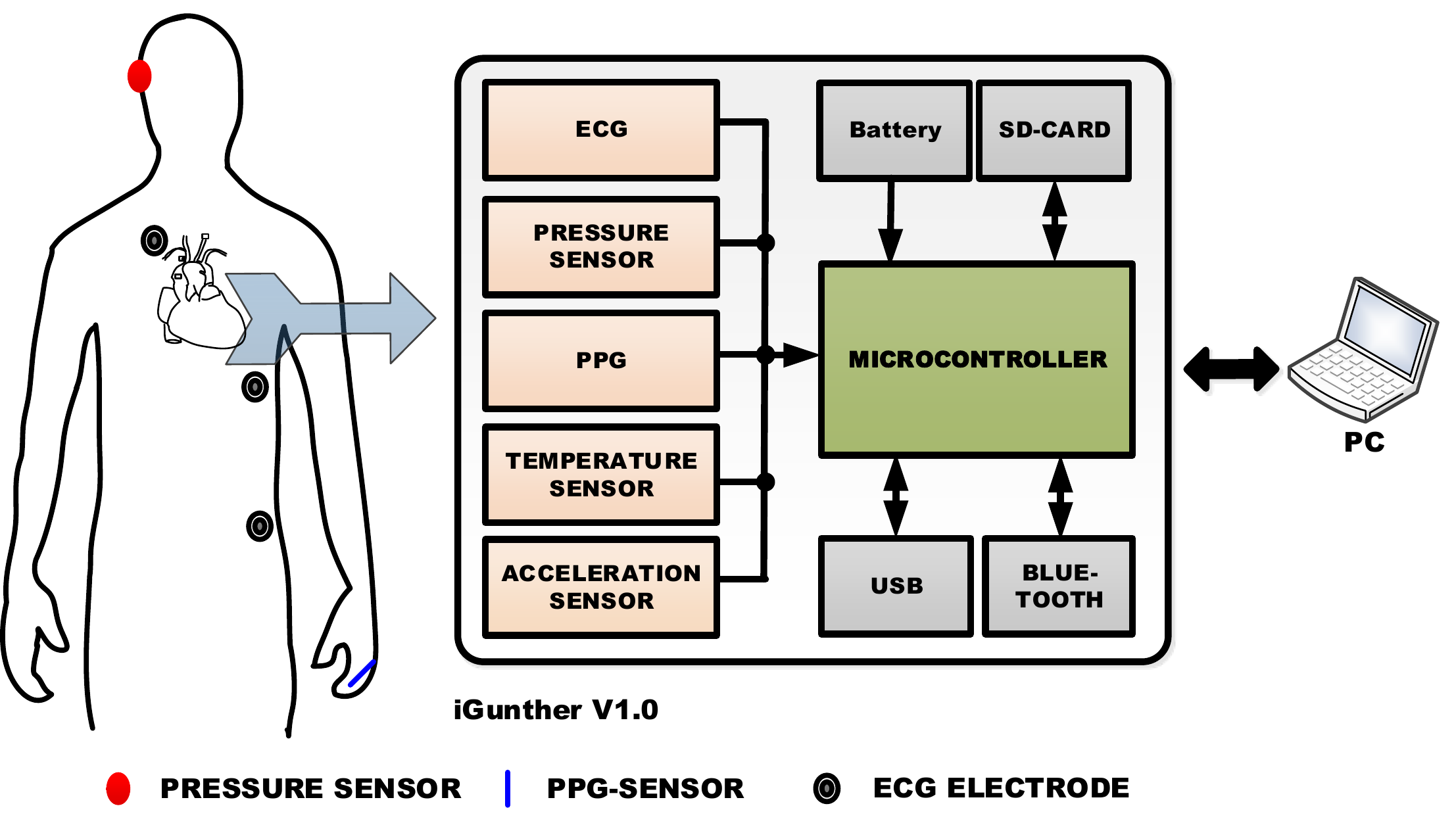}
	\caption{Block diagram of the microcontroller based measurement system.}
	\label{fig:SystemBlockDiagram}
\end{figure}

Fig. \ref{fig:SystemBlockDiagram} shows a block diagram of the developed system. The microcontroller based embedded system acquires all measurement data from the subject and transmit the measured data to the host PC for further analysis and display. The data can be transmitted via Bluetooth / USB or can be stored on the micro SD-card. To allow a higher degree of flexibility the system can be battery powered, too.

\subsection{ECG Module}
The developed ECG module consists of an Instrumentation Amplifier (INA) with provisions for a Driven Right Leg (DRL) and a driven shield circuit, a base line wandering rejection circuit, as well as an optional $50~Hz$ notch filter to mitigate power-line noise. The ECG module is able to measure the ECG in a frequency range of $50~mHz$ to $150~Hz$. The analog ECG signal is digitalized via the internal $12~bit$ Analog to Digital Converter (ADC) of the microcontroller.

\subsection{Photoplethysmography (PPG) Module}
The implemented PPG circuit is designed for reflective mode and is based on two LED with wavelengths of $660~nm$ and $940~nm$. The LED are wired in a way, that only one LED can be switched on at a time. The reflected light will be sensed by a photo diode, which is connected to a transimpedance amplifier (OPA2380 from Texas Instruments). Based on the voltage present at the output of the amplifier, digitalized via the ADC of the microcontroller, the LED current is regulated via the microcontroller's Digital to Analog Converter (DAC) to keep the output voltage of the transimpedance amplifier constant. Afterwards the analog PPG signal is high-pass filtered, to remove the DC-offset due to the dark current of the photo diode, with a servo feedback high-pass filter with a cutoff frequency of $\approx 15~mHz$. A subsequent 5th-order Sallen-Key low-pass filter with a cutoff frequency of $\approx 30~Hz$ is used to mitigate $100~Hz$ flickering of the ambient light. The cutoff frequencies are chosen to allow a distortion free investigation of the expected PPG waveform, with its spectral components ranged mainly between a few $mHz$ up to $\approx 20~Hz$. The band-pass filtered analog PPG signal is afterwards digitalized via another channel of the microcontrollers ADC.

\subsection{Pressure Measurement}
The pressure sensors (HCE-M010DBE8P3 from Fist Sensor AG) have a calibrated and compensated pressure measurement range of $\pm 1000~Pa$ ($\pm 10~mbar$). The pressure sensor supports up to $1~kSPS$ with $14~bit$ precision, which leads to a theoretic resolution of $0.122~Pa$. The temperature sensors are read out via an Serial Peripheral Interface (SPI) which is connected to the microcontroller. 

\subsection{Acceleration Measurement}
The acceleration measurement is based on a LIS3DH (ST Microelectronics) acceleration sensor. The sensor can measure in three directions with full scale ranges from $\pm 2~g$ up to $\pm 16~g$ with $16~bit$ precision per direction at a sample rate from $1~Hz$ to $5~kHz$. Furthermore the sensor has a build-in functionality for free-fall detection and is connected to the microcontroller via the SPI interface.

\subsection{Temperature Measurement}
The temperature measurement is based on a NTC sensor with dimensions of $1.0~mm \times 0.5~mm$. The sensor is therefore small enough to be located inside the ear. The NTC resistance is evaluated with a quarter Wheatstone bridge and will be calibrated to $37~^\circ C$ for optimal sensitivity. The output signal of the bridge is amplified and digitalized with the internal ADC of the microcontroller.

\subsection{Housing and Battery}
\begin{figure}[!h]
	\centering
	\includegraphics[width=0.32\textwidth]{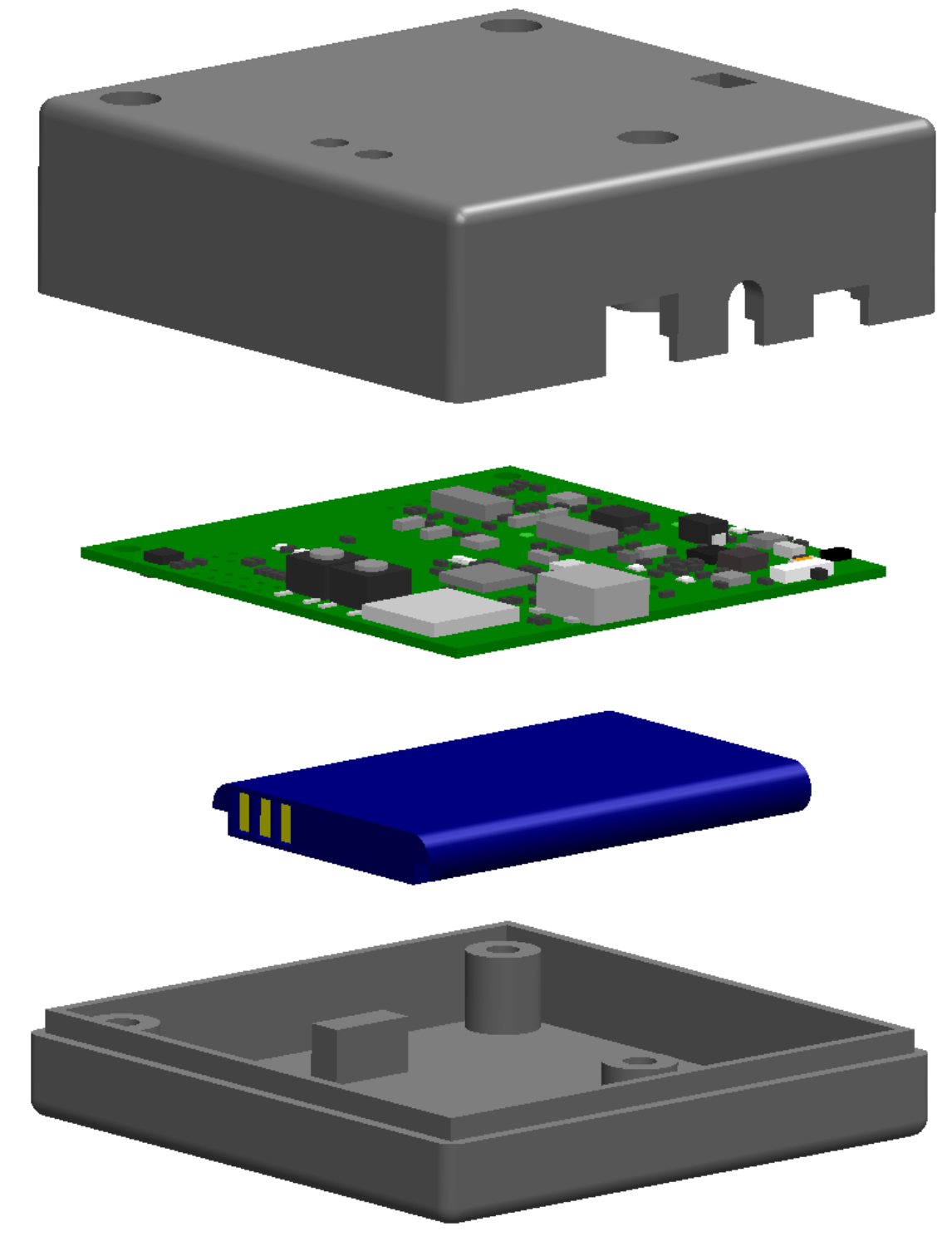}
	\caption{Exploded assembly drawing of the housing of the measurement system inclusive the Printed Circuit Board (PCB) and the used battery.}
	\label{fig:Housing}
\end{figure}

The housing for the measurement system was designed in SolidWorks (Dassault Systems) and manufactured with a 3d-printer (MakerBot Replicator 2X). The housing consists of ABS plastic and has a size of $71.5~mm \times 71.5~mm \times 38~mm$. Fig. \ref{fig:Housing} shows a exploded assembly drawing of the housing incl. Printed Circuit Board (PCB) and the used battery. The li-ion battery has $3.7~V$ by $1.25~Ah$ with dimensions of about $53~mm \times 34~mm \times 5.5~mm$ and a weight of about $23~g$. The expected system run-time with the battery is about 8 hours.

\subsection{Microcontroller System and Communication Interfaces}
The microcontroller (ATxMega128A4U from Atmel) acquires the measurement data from the different sensors and is also responsible for the data transmission to the host PC via Bluetooth or USB and for the optional storage of the measurement data on the micro SD-card. While USB communication is realized via the internal USB stack of the microcontroller, Bluetooth is implemented with a commercially available module (RN42-I/RM from Roving Networks). The Bluetooth module is certified according to Bluetooth V2.1 incl. Enhanced Data Rate (EDR) mode and supports master and slave mode with up to $300~kbps$. The interfacing to the microcontroller is implemented as asynchronous serial connection.

The system is externally supplied either via USB or via the system battery. Whereas the digital components operate with $+3.3~V$ only, the analog components are supplied with $\pm 3.3~V$. For electrical safety considerations it is recommended to operate the system in wireless or SD-card storage mode while running from the system battery.

\subsection{Manufactured System}
\begin{figure}[!h]
	\centering
	\includegraphics[width=0.42\textwidth]{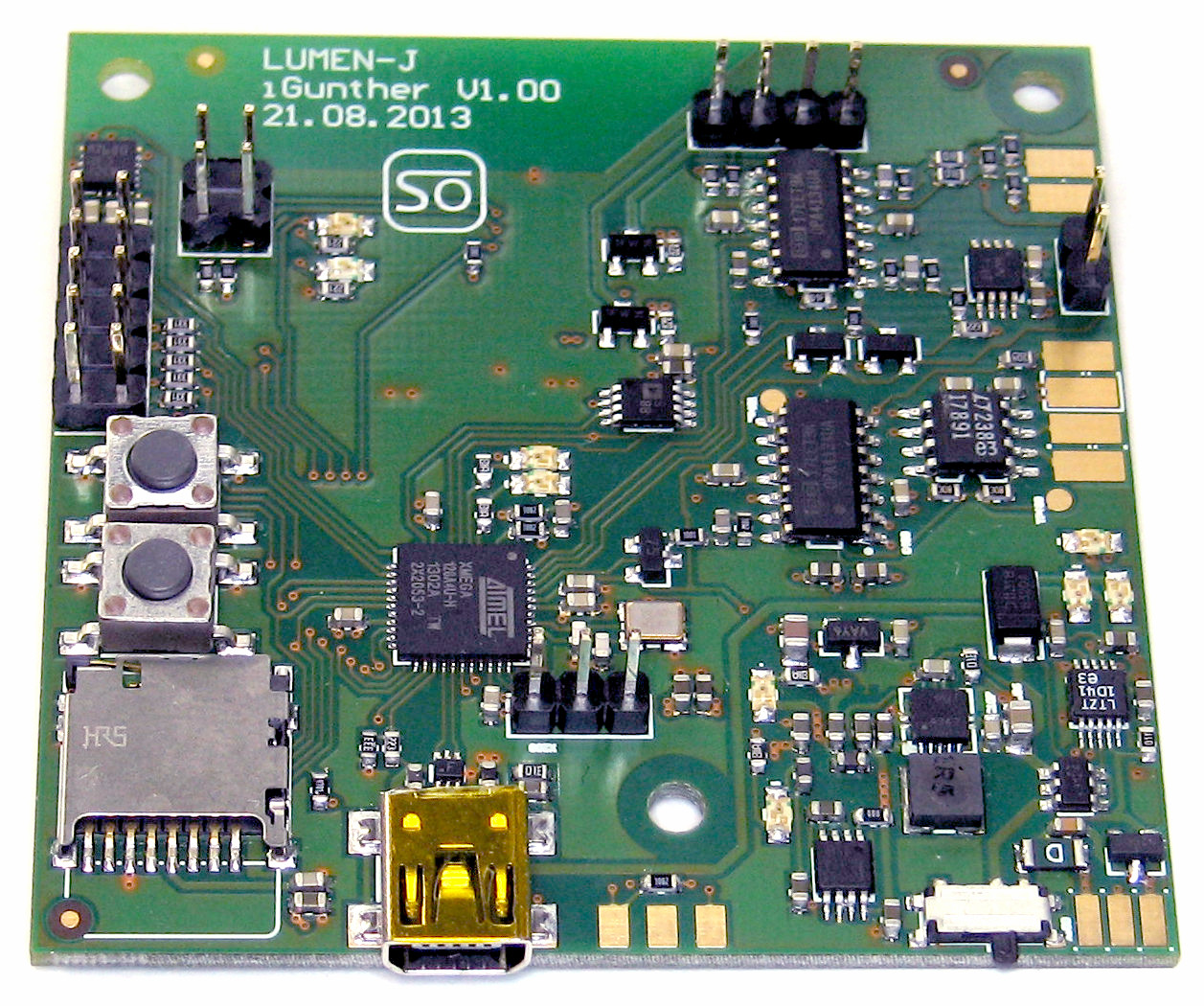}
	\caption{Manufactured and populated Printed Circuit Board (PCB) of the developed measurement system. The PCB has dimensions of $60~mm \times 60~mm$ and contains about 200 components.}
	\label{fig:PCB}
\end{figure}

Fig. \ref{fig:PCB} shows a photograph of the manufactured and populated Printed Circuit Board (PCB) of the developed measurement system. The PCB has a size of about $60~mm \times 60~mm$ and contains about 200 components.

\subsection{Software}
The firmware of the developed measurement system is based on the Atmel Software Framework (ASF) and is written in C language.

\begin{figure}[!h]
	\centering
	\includegraphics[width=0.43\textwidth]{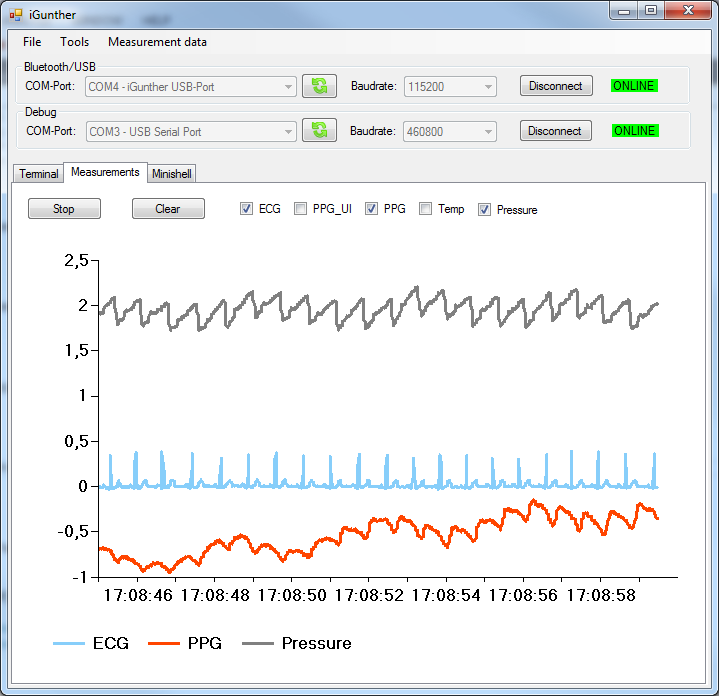}
	\caption{Screenshot of the developed C\# based PC interface software.}
	\label{fig:SW}
\end{figure}

Tbe PC interface software is written in C\# language and is based on the .NET framework from Microsoft. It is able to display the measured waveforms in real-time with a latency of about $50~ms$ and can be used to configure the embedded measurement systems in terms of active channels and sample rates. With the interface software, waveforms can be recorded and exported to MathWorks MATLAB or Excel. Fig. \ref{fig:SW} shows an image of the interface software. Visible is additional to the Graphical User Interface (GUI) the pressure inside the ear, the ECG, as well as a PPG, acquired with the developed system. It can be seen, that the waveforms have the expected typical shapes. Based on the recorded waveforms calculations of the PAT and PVW values based on PPG, ECG and the pressure signal are possible.

\section{RESULTS AND DISCUSSION}

Table \ref{table:measurements} shows preliminary measurement results acquired over a period of 180 seconds on three healthy subjects. For simplicity reasons only measurements on the left ear were taken and the PWV was calculated while assuming a PEP of $0~s$. The PPG sensor was placed around incisura and tragus targeting the arteria temporalis superficialis and arteria auriculares anteriors.

The acquired measurements have a sample rate of $1~kHz$ and are filtered via wavelet decomposition in MathWorks MATLAB. The filtered signals are fed to an automated peak detection and are referenced to the ECG for evaluating the PAT. Together with the measured distance ($\Delta x$) the PWV is calculated according to (\ref{equ:PWV}).

The PAT derived from the pressure measurements are in a range of $100~ms$ to $200~ms$ and the PAT derived from the PPG measurements are in a range of $200~ms$ to $240~ms$. The results are reasonable taking into account that usually reported PAT values on extremities are in the range of $> 240~ms$ \cite{Comparisonofbilateralpulsearrivaltimebeforeandafterinducedvasodilationbyaxillaryblock} and that \cite{kaufmann:Inearpulsewavemeasurements} reported PAT values for in-ear pressure measurements around 100 ms. 
\begin{table}[!h]
	\caption{{\normalsize \textbf{Measurements}}}
	\label{table:measurements}
	\centering
	\begin{tabular}{l l l l }
	   & {\small \textbf{Subject 1}} & {\small \textbf{Subject 2}} & {\small \textbf{Subject 3}}\\
	\hline
	{\small \textbf{Age}}							&	$25$				&	$29$					& $33$\\
	{\small \textbf{Sex}}			    			&	male				&	male					& female\\
	{\small \textbf{Height}}		    			&	$170~cm$			&	$188~cm$				& $158~cm$\\
	{\small \textbf{PAT length ($\Delta x$)}}		&	$27~cm$				&	$30~cm$					& $26~cm$\\
	{\small \textbf{Weight}}		    			&	$68~kg$				&	$134~kg$				& $52~kg$\\
	{\small \textbf{Heart Rate}}					&	$68~BPM$			&	$72~BPM$				& $70~BPM$\\
	{\small \textbf{Mean $PAT_{peak,~PPG}$}}		&	$207~ms$			&	$212~ms$				& $235~ms$\\
	{\small \textbf{Mean $PAT_{peak,~Pressure}$}}	&	$114~ms$			&	$171~ms$				& $123~ms$\\
	{\small \textbf{Mean $PWV_{PPG}$}}				&	$1.3~m/s$			&	$1.4~m/s$				& $1.1~m/s$\\
	{\small \textbf{Mean $PWV_{Pressure}$}}			&	$2.4~m/s$			&	$1.8~m/s$				& $2.1~m/s$\\
	\hline
	\end{tabular}
\end{table}

\section{CONCLUSION}
A prototype of an in-ear pulse wave measurement system was developed and tested. The results are very promising and showing a good performance. In future the prototype has to be miniaturized and the interface software must be enhanced by the possibilities to measure PAT and PWV, as well as the heart rate directly. Furthermore a second PPG channel for measurements on the extremities should be added to have an additional reference measurement side for comparisons.

\section*{Acknowledgment}
This publication is a result of the ongoing research within the LUMEN research group, which is funded by the German Federal Ministry of
Education and Research (BMBF, FKZ 13EZ1140A/B). LUMEN is a joint research project of L\"ubeck University of Applied Sciences and University of L\"ubeck and represents an own branch of the Graduate School of University of L\"ubeck.

Furthermore the authors would like to thank Texas Instruments and Linear Technology for their support in terms of free samples during the development process.

\end{document}